\documentclass[aip,apl,reprint]{revtex4-1}

\usepackage{graphicx}
\usepackage{booktabs}
\usepackage{bm}
\usepackage{longtable}
\usepackage{amsmath}
\usepackage{dcolumn}
\usepackage{placeins}
\usepackage{expdlist}

\usepackage{siunitx}

\begin{document}


\title{Non-destructive structural imaging of steelwork with atomic magnetometers}

\author{P. Bevington} 
\affiliation{National Physical Laboratory, Hampton Road, Teddington, TW11 0LW, United Kingdom}
\author{R. Gartman} 
\affiliation{National Physical Laboratory, Hampton Road, Teddington, TW11 0LW, United Kingdom}
\author{W. Chalupczak}
\affiliation{National Physical Laboratory, Hampton Road, Teddington, TW11 0LW, United Kingdom}
\author{C. Deans}
\affiliation{Department of Physics and Astronomy, University College London, London WC1E 6BT, United Kingdom}
\author{L. Marmugi}
\affiliation{Department of Physics and Astronomy, University College London, London WC1E 6BT, United Kingdom}
\author{F. Renzoni}
\affiliation{Department of Physics and Astronomy, University College London, London WC1E 6BT, United Kingdom}

\date{\today}

\begin{abstract}
We demonstrate imaging of ferromagnetic carbon steel samples and we detect the thinning of their profile with a sensitivity of $\SI{0.1}{\milli\meter}$ using a Cs radio-frequency atomic magnetometer. Images are obtained at room temperature, in magnetically unscreened environments. By using a dedicated arrangement of the setup and active compensation of background fields, the magnetic disturbance created by the samples' magnetization is compensated. Proof-of-concept demonstrations of non-destructive structural evaluation in the presence of concealing conductive barriers are also provided. Relevant impact for steelwork inspection and health and usage monitoring without disruption of operation is envisaged, with direct benefit for industry, from welding in construction, to pipelines inspection and corrosion under insulation in the energy sector. 
\vskip 20pt

\begin{center}
This is a preprint version of the article appeared in Applied Physics Letters:\\
P. Bevington, R. Gartman, W. Chalupczak, C. Deans, L. Marmugi, F. Renzoni, \\Appl. Phys. Lett. \textbf{113}, 063503 (2018) DOI: \href{https://doi.org/10.1063/1.5042033}{10.1063/1.5042033}.
\end{center}
\end{abstract}


\maketitle

Non-destructive inspection of pipelines, vessels, and structural steelwork is an important open challenge for various industry sectors. Anomalies or material fatigue can have severe consequences. 
For example, in manufacturing and construction the quality of assemblies and welding is critical, and often requires the use of dangerous and expensive X-ray scans. In health and usage monitoring systems (HUMS) timely and non-invasive identification of structural damages and fatigue is a primary target. In the energy sector spillage has economical as well as environmental impacts.
Specifically, corrosion under insulation (CUI) accounts for 60$\%$ of pipe leaks, causing significant losses due to unscheduled downtime and maintenance. This is further exacerbated by the presence of thick insulating layers which conceals the corroded part. Corrosion-related costs in industry can exceed \$270bn/year \cite{koch2001}.

A wealth of technologies have been proposed for assessing the structural integrity of steelwork and pipelines. These include ultrasound tomography, microwave sensing, acoustic emission, and electrochemical impedance spectroscopy \cite{OGTC2016}. Such techniques are invasive, i.e. require direct access to the tested surface. 
Eddy current testing is a widely used non-destructive evaluation (NDE) method to identify cracks and fatigue-related damage in metallic structures \cite{Auld1999, Griffiths2001, Perez2004, Sophian2017}, as well as to detect impurities in fluids. It relies on the generation of eddy currents by an oscillating magnetic field (the primary field, referred to as the ``rf field'') in the object of interest and on the detection of the magnetic field produced by those eddy currents (the secondary field). Position-resolved measurements then allow the reconstruction of the image of the object in the form of a conductivity map. In the case of ferromagnetic metallic objects, which have a relatively high permeability and low conductivity, the secondary field originates from an oscillating local magnetisation induced by the primary field and not from eddy currents.

Here, we present imaging of ferromagnetic samples, using an ultra-sensitive atomic magnetometer \cite{Wickenbrock2014, Deans2016, Wickenbrock2016}, with active magnetic field compensation system \cite{Deans2017}, and a dedicated measurement geometry, suitable for industrial monitoring.  In particular, we demonstrate imaging and measurement of changes in the thickness of pipeline-grade carbon steel. This measurement, accepted by industry as a benchmark \cite{Winnik2015}, represents a proof-of-concept demonstration of the relevance of the atomic magnetometer technology in steelwork NDE and CUI detection.

\begin{figure}[htbp]
\includegraphics[width=\columnwidth]{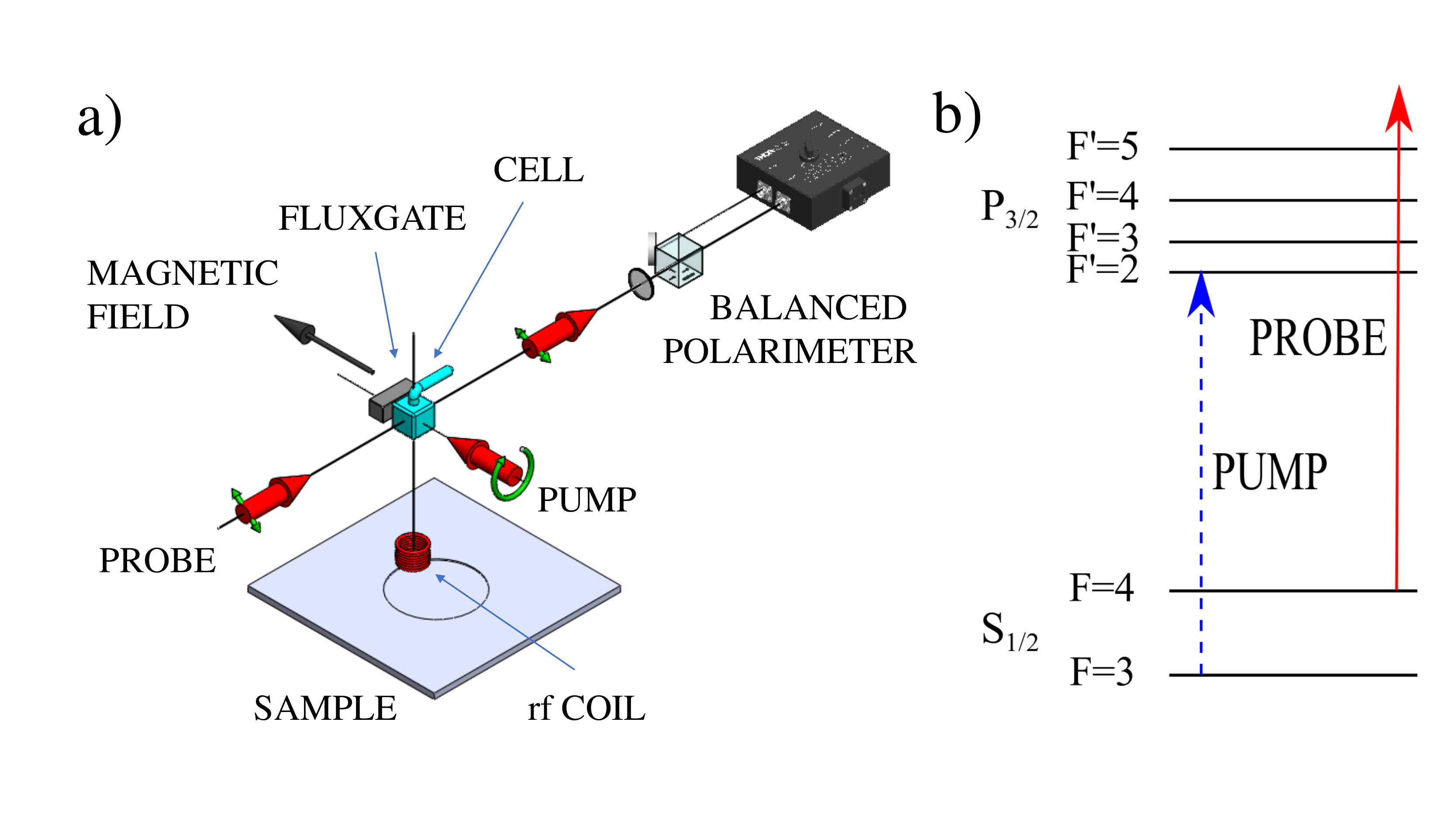}
\caption{(a) Main components of the experimental setup. (b)  Caesium $6\,^2$S$_{1/2}$ F=3$\rightarrow{}6\,^2$P$_{3/2}$ F'=2 transition (D2 line, 852 nm) energy structure (detunings of the pump and probe laser beams marked with dashed blue and solid red lines).}\label{fig:Setup}
\end{figure}

Figure~\ref{fig:Setup} shows the key components of the experimental setup. Details of the atomic magnetometer configuration are described in \cite{Chalupczak2012}. Here we only recall key elements. Detection of the secondary field is performed with a  $\SI{1}{\centi\meter\cubed}$ paraffin coated glass cell containing room temperature cesium vapour (atomic density $n_{\text{Cs}}=\SI{3.3e10}{\per\centi\meter\cubed})$.  Atoms are pumped with a circularly polarized pump laser beam ($\SI{377}{\micro\watt}$), frequency locked to the cesium $6\,^2$S$_{1/2}$ F=3$\rightarrow{}6\,^2$P$_{3/2}$ F'=2 transition (D2 line, $\SI{852}{\nano\meter}$) propagating along the bias magnetic field.  Coherent atomic spin precession is driven by the rf field. The superposition of the primary and the secondary fields alters this motion, which is probed with a linearly polarized probe laser beam propagating orthogonally to the bias magnetic field. The probe beam ($\SI{30}{\micro\watt}$) is phase-offset-locked to the pump beam, bringing it $\SI{580}{\mega\hertz}$ blue shifted from the $6\,^2$S$_{1/2}$ F=4$\rightarrow{}6\,^2$P$_{3/2}$ F'=5 transition (D2 line, $\SI{852}{\nano\meter}$). Faraday rotation is detected with a balanced polarimeter, whose signal is then processed by a lock-in amplifier referenced to the phase of the rf field. The rf coil axis is orthogonal to both the pump and probe beam.

This work was carried out in a magnetically unshielded environment. Three pairs of mutually orthogonal square Helmholtz coils (largest coil length $\SI{1}{\meter}$) are used for active and passive compensation of the ambient magnetic field and for adjusting the direction and strength of the bias magnetic field. 

\begin{figure}[htbp]
\includegraphics[width=\columnwidth]{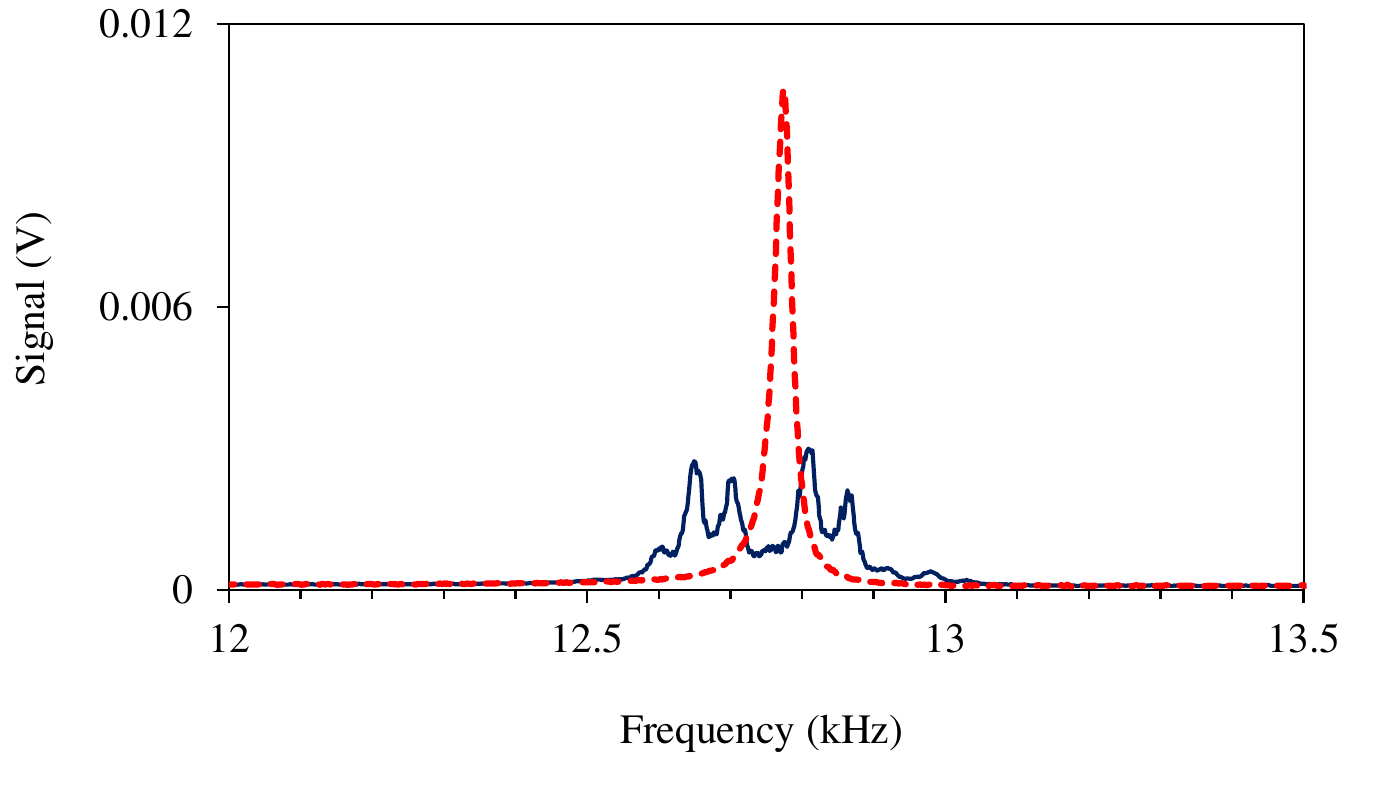}
\caption{ rf spectra with (dashed red line) and without (solid blue line) active stabilisation. The plots show the amplitude output of the lock-in amplifier ($\SI{10}{\milli\second}$ time constant), while the rf frequency in ramped at $\SI{25}{\hertz\per\second}$.}\label{fig:Spectrum}
\end{figure}

Figure~\ref{fig:Spectrum} shows the rf spectra generated by atoms in the F=4 ground states with and without the active compensation. Significant broadening of the resonance profile was observed without active compensation, due to slow frequency drifts of the environmental magnetic field and sidebands that correspond to $\SI{50}{\hertz}$ noise produced by electronic devices. To perform active compensation we use a commercial fluxgate (Bartington Mag690) located next to the vapour cell and three PID units (SRS 960). With passive and active field compensation the linewidth of the rf spectral profile is approximately $\SI{30}{\hertz}$. The bandwidth of the three independent servo loops spans from DC to $\SI{3}{\kilo\hertz}$. We measured a reduction of 10 times for the dominant $\SI{50}{\hertz}$ noise. See also \cite{Belfi2010} for an alternative approach to spurious magnetic field compensation.
The small size of the atomic cell provides partial immunity to ambient field gradients. In this way, gradient compensation is not necessary. The ambient magnetic field gradient is estimated to be in the order of  $\SI{200}{\nano\tesla\per\centi\meter}$ which corresponds to approximately $\SI{20}{\hertz}$ of broadening \cite{Pustelny2004}, assuming it is all directed along the bias magnetic field.

Eddy currents are excited by the same rf coil (1000 turns of $\SI{0.2}{\milli\meter}$ diameter copper wire, height $\SI{10}{\milli\meter}$,  $\SI{2}{\milli\meter}$ and $\SI{4}{\milli\meter}$ inner and outer diameters) which drives the atomic magnetometer. The coil is placed $\SI{2}{\milli\meter}$ from the object (coil lift-off). This arrangement generates a larger density of eddy currents due to the small distance between the rf coil and the sample. The sample plate ($150\times\SI{150}{\milli\meter\squared}$) is placed on a 2D translation stage actuated by two computer controlled stepper motors with $\SI{0.184}{\milli\meter}$ positioning precision. 

The samples used in this letter are made of $\SI{6}{\milli\meter}$ thick carbon steel, a type commonly used in the energy sector. Contrary to previous works \cite{Deans2017}, in this case the imaging target is a ferromagnetic material. As such, its magnetic signatures cannot be ruled as mere background, nor can they be considered unchanged or predictable among different measurements. Carbon steel has a macroscopic non-zero magnetic moment that is imprinted during molding, and is changed by physical stresses and further treatment processes. Unpredictable variations in magnetic moment along the surface of the sample create strong field gradients. To reduce the impact of such anomalies the sample is located approximately $\SI{300}{\milli\meter}$ from the atomic sensor. Any residual DC magnetic field created by the ferromagnetic object at the sensor's location is automatically zeroed by our field compensation system. Non-ideal full field compensation results from a non-zero distance between the fluxgate head and vapor cell. This could be improved by implementation of the compensation scheme discussed in \cite{Belfi2010}. In our configuration the observed rf resonance profile frequency shift across all six samples is between $\SI{210}{\hertz}$ and $\SI{850}{\hertz}$ in a $64\times\SI{64}{\milli\meter\squared}$  scan range.

\begin{figure}[htbp]
\includegraphics[width=\columnwidth]{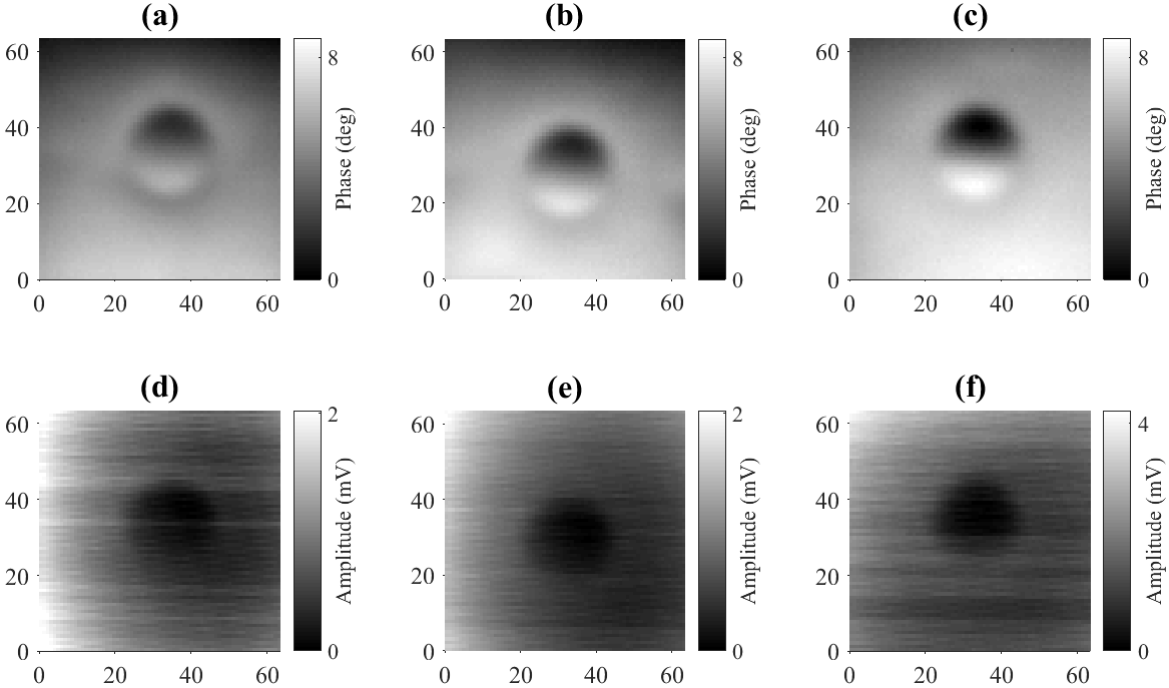}
\caption{Phase (a, b, c) and amplitude (d, e, f) change in rf signal generated by the scans of $64\times\SI{64}{\milli\meter\squared}$ area of carbon steel plate with a defect ($\SI{24.5}{\milli\meter}$ diameter) in a form of recess - $20\%$ (a, d), $40\%$ (b, e) and $60\%$ (c, f) of the plate thickness. The images have been recorded at $\SI{12.6}{\kilo\hertz}$.}\label{fig:Image}
\end{figure}

Figure  \ref{fig:Image} shows results of the scans of $64\times\SI{64}{\milli\meter\squared}$ area of carbon steel plate with a defect ($\SI{24.5}{\milli\meter}$ diameter) in the form of a recess. Three pairs of images represent measurements with three recess depths $20\%$ (a, d), $40\%$ (b, e) and $60\%$ (c, f) of the plate thickness. These mimic local thinning due to structural anomalies such as different levels of corrosion, or fatigue. 
Each pixel of the image represents the peak-amplitude [Fig. 3(d-f)] and the corresponding phase at resonance [Fig. 3 (a-c)] of the rf profile recorded by scanning the frequency through the magnetic resonance. These were determined by automatically fitting the amplitude and the phase of the resonance curves (see Fig.~\ref{fig:Spectrum} for an example of amplitude curve).
Both resonance amplitude and phase reveal the presence of the recess. It is worth pointing out that the amplitude of the rf field is reduced due to the presence of a recess [Fig.~\ref{fig:Image} (d-f)]. We observed that the amplitude change observed in a non-magnetic, highly-conductive sample (aluminium) has the opposite sign. This indicates that the signals in these two cases have a different origin. In the case of a carbon steel sample, we measure effects created by the AC samples' magnetization induced by the rf field, rather than eddy current induction, whereas in the case of the highly conductive and non-magnetic aluminium the main contribution is produced exclusively by eddy currents. The magnetisation induced in the sample is in the same direction as the primary field and hence, the presence of the recess lowers the value of total field. 
The changes in rf signal phase that produce a `dispersive'-like profile could be intuitively understood in terms of the modifications of the resulting secondary field's symmetry and orientation. In the case of a uniform sample surface, the secondary field is parallel to the primary field, i.e. orthogonal to the surface. However, the presence of inhomogeneities breaks the symmetry and changes the orientation of the secondary field. The asymmetry of the magnetization is reversed on opposite side of the recess.

We have intentionally chosen dimensions of the plate significantly bigger than the diameter of the defect so that the image is not disturbed by the signal generated by the edges of the plate. We operate our measurement at $\SI{12.6}{\kilo\hertz}$, which corresponds to a skin depth estimated to be $\SI{0.18}{\milli\meter}$. The rf field penetrates much deeper in the sample, although with an exponentially decreasing amplitude \cite{Stucky1990}. 

The relative orientation of the sample with respect to the sensor does not prevent imaging of the defect: similar images, although with smaller contrast, were obtained with the recess facing the sensor and with the recess opposite to the sensor. The latter mimics damage or corrosion in the inner face of a steel pipeline.

In the following, we discuss a number of properties of the recorded images and demonstrate that our imaging system is capable of discriminating different levels of thinning of the sample. 
In the course of systematic measurements with metallic samples we have observed that the edges of the recess generate `dispersive'-like profile in rf signal phase with `linewidth' of  about $\SI{20}{\milli\meter}$. We ascribe such behavior within the recess boundaries 
to the interplay between the size of the probe (rf coil) and the defect profile (recess) \cite{Auld1999}. This indicates that   - for the given coil size - the contrast observed in measurement is limited by the defect size, which is confirmed by the smaller phase variation recorded for samples with smaller recesses ($\SI{12}{\milli\meter}$ diameter) in the same conditions.

\begin{figure}[htbp]
\includegraphics[width=\columnwidth]{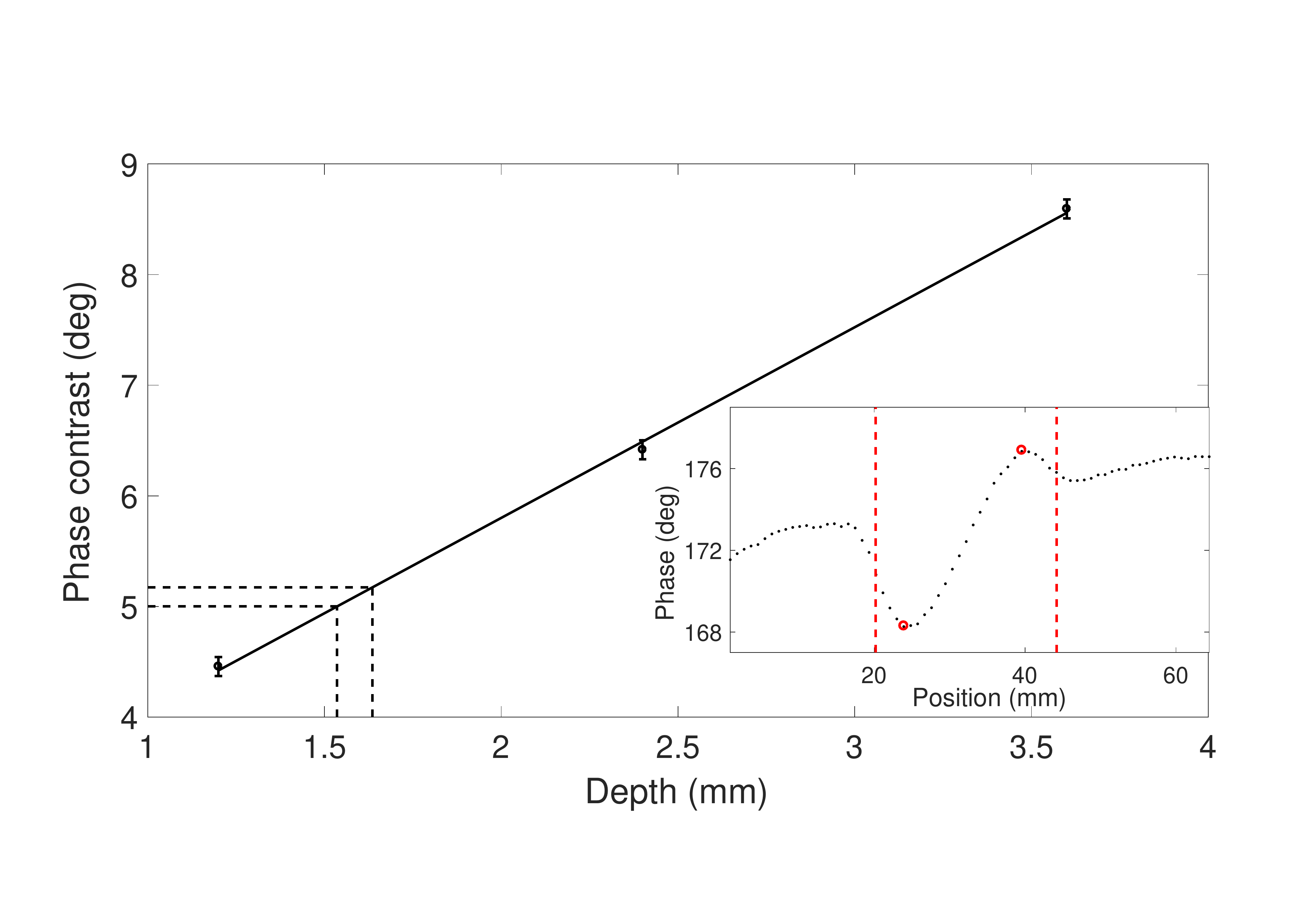}
\caption{Phase contrast as a function of the recess depth. Error bars represent uncertainty of the rf spectrum fit results and indicate the thickness measurement resolution at the level of $\SI{0.1}{\milli\meter}$. Inset: vertical cross sections across the image shown in Fig.~\ref{fig:Image}(c). Red points mark the maximum and minimum phase recorded within the recess, and used for the calculation of phase contrast. Dashed red lines mark the actual position and size of the recess.}\label{fig:crosssections}
\end{figure}

In order to quantify the results of the observations, we introduce the phase contrast, which we define as the difference between the maximum and minimum phase within the recess boundaries. Figure~\ref{fig:crosssections} shows the dependence of the phase contrast on the depth of the defect. It demonstrates that this measurement is able to resolve $\SI{0.1}{\milli\meter}$ change in sample thickness. This is indicated by the dashed lines in Fig.~\ref{fig:crosssections} that map the phase contrast error to the corresponding depth uncertainty. Similar considerations demonstrate thickness resolution of  $\SI{0.6}{\milli\meter}$ for the recesses with $\SI{12}{\milli\meter}$ diameter.

The inset of Fig.~\ref{fig:crosssections} shows a cross-section of the phase images from Fig.~\ref{fig:Image}(c). Clear detection of the recess is shown, with measured size comparable within $\SI{5}{\milli\meter}$ to the recess actual dimensions. Thus, the amplitude of the  phase change is related to the depth of the recess, and thus enables local thickness estimation, while its extension on the plate plane allows to determine the area of the recess.

\begin{figure}[htbp]
\includegraphics[width=\columnwidth]{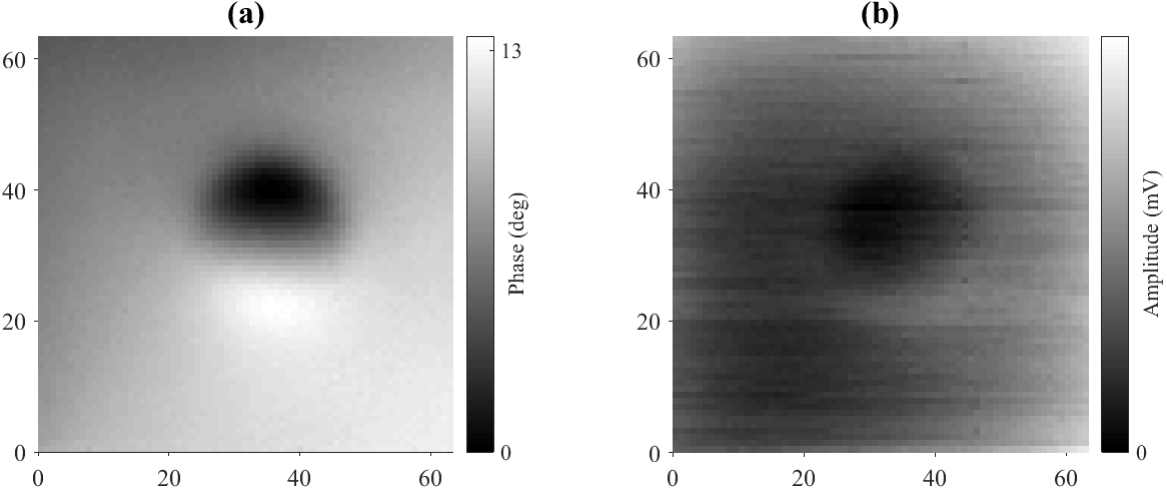}
\caption{Phase (a) and amplitude (b) change in rf signal generated by the scans of $64\times\SI{64}{\milli\meter\squared}$ area of carbon steel plate with a $\SI{24.5}{\milli\meter}$ diameter recess, $60\%$ of the plate thickness. The carbon steel plate is covered by an Al sheet ($\SI{0.5}{\milli\meter}$ thickness.  The images have been recorded at $\SI{12.6}{\kilo\hertz}$.}\label{fig:CUI}
\end{figure}

To simulate the realistic situation of barriers concealing the region of interest (e.g. insulating layers, support structures, etc), we introduce an aluminium sheet to mimic the the worst case scenario of conductive insulation materials. Figure~\ref{fig:CUI} shows results of the scan of $64\times\SI{64}{\milli\meter\squared}$ area of carbon steel plate with $\SI{24.5}{\milli\meter}$ diameter recess, $\SI{3.6}{\milli\meter}$ deep.  A $\SI{0.5}{\milli\meter}$ thick aluminium sheet is placed on top of the sample. The image has been recorded at $\SI{12.6}{\kilo\hertz}$, where the skin depth for aluminium is $\SI{0.7}{\milli\meter}$. In this case, the concealing layer and the steel sample are in electrical contact. Analogous images were obtained when the Al sheet and the carbon steel sample are not in electrical contact.
This demonstrates the ability of imaging through a concealing barrier, of direct relevance to pipeline monitoring in the energy sector.

In conclusion, we have demonstrated the relevance of eddy current imaging with atomic magnetometers for NDE of steelwork in industrial monitoring and HUMS, for example for the detection of corrosion under insulation. Because of the high sensitivity of the Cs rf atomic magnetometer we were able to detect changes in thickness of the carbon steel samples with $\SI{0.1}{\milli\meter}$ resolution even with the system operating at the relatively high frequency of $\SI{12}{\kilo\hertz}$. Ongoing work is focused on identification of the factors limiting spatial resolution of the measurements.

\begin{acknowledgements}
This work was funded by the Innovate UK Energy Game Changer programme (IUK 132437). 
\end{acknowledgements}

\end{document}